\documentclass[12pt]{article}
\usepackage{amsmath,amssymb,dsfont}
\usepackage{graphicx}

\newtheorem{theorem}{Theorem}

\newtheorem{definition}[theorem]{Definition}

\begin{document}
\title{Optimization of distributed EPR entanglement generated between two Gaussian fields by the modified steepest descent method}
\author{Zhan Shi and Hendra I. Nurdin
\thanks{
Z. Shi and H. I. Nurdin are with School of Electrical Engineering and 
Telecommunications,  UNSW Australia,  
Sydney NSW 2052, Australia (e-mail: zhan.shi@student.unsw.edu.au,  h.nurdin@unsw.edu.au).} 
}
\maketitle

\begin{abstract}
Recent theoretical investigations on quantum coherent feedback networks have found that with the same pump power, the Einstein-Podolski-Rosen (EPR)-like entanglement generated via a dual nondegenerate optical parametric amplifier (NOPA) system placed in a certain coherent feedback loop is stronger than the EPR-like entangled pairs produced by a single NOPA. In this paper, we present a linear quantum system consisting of two NOPAs and a static linear passive network of optical devices. The network has six inputs and six outputs, among which four outputs and four inputs are connected in a coherent feedback loop with the two NOPAs. This passive network is represented by a $6 \times 6$ complex unitary matrix. A modified steepest descent method is used to find a passive complex unitary matrix at which the entanglement of this dual-NOPA network is locally maximized. Here we choose the matrix corresponding to a dual-NOPA coherent feedback network from our previous work as a starting point for the modified steepest descent algorithm. By decomposing the unitary matrix obtained by the algorithm as the product of so-called two-level unitary matrices, we find an optimized configuration in which the complex matrix is realized by a static optical network made of beam splitters.
\end{abstract}

\section{Introduction}
\label{sec:intro}
In recent years, research related to the Einstein-Podolski-Rosen (EPR) entanglement in continuous-variable quantum information processing has become increasingly vital since it can be shared by two distant communicating parties and is used as the crucial resource for important applications such as quantum teleportation and superdense coding \cite{BSLR}. Compared to discrete variable entangled states, continuous variable entanglement such as EPR entanglement \cite{BSLR} is generated efficiently by a pair of squeezed light beams and utilized with expeditiousness in measurement of quantum states which is a critical step in quantum communication protocols \cite{BL2005, Weedbrook2012}.

A device that is used to produce EPR-like entangled states is a nondegenerate optical parametric amplifier (NOPA), which contains a cavity with a $\chi^{(2)}$ nonlinear crystal inside. Via a strong undepleted coherent beam pumped to the crystal, two ingoing signals in vacuum state interact with two modes of the cavity separately, and generate two output beams which are squeezed in quadrature-phase amplitudes and considered as EPR entanglement \cite{Ou1992}. EPR entanglement between the two outgoing fields is measured by the two-mode squeezing spectra of the fields. A strong EPR entanglement is denoted by a high degree of two-mode squeezing. As a quantum system in reality is sensitive to its external environment, it undergoes unwanted interaction with an external electromagnetic field as a ``heat bath''  \cite{GardinerBook}.  Therefore the strength of EPR entanglement generated by such an open system can be degraded due to transmission losses and decoherence, which leads to limited communication distance. Thus methods to enhance EPR entanglement are of interest to improve the quality of quantum communication.

\begin{figure}[htbp]
\begin{center}
\includegraphics[scale=0.28]{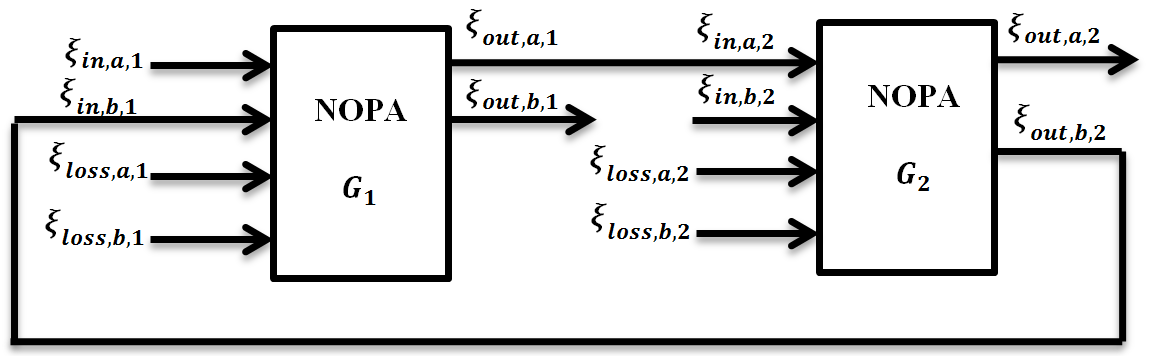}
\caption{The dual-NOPA coherent feedback network.}\label{fig:dual-NOPA-cfb}
\end{center}
\end{figure}
\begin{figure}[htbp]
\begin{center}
\includegraphics[scale=0.27]{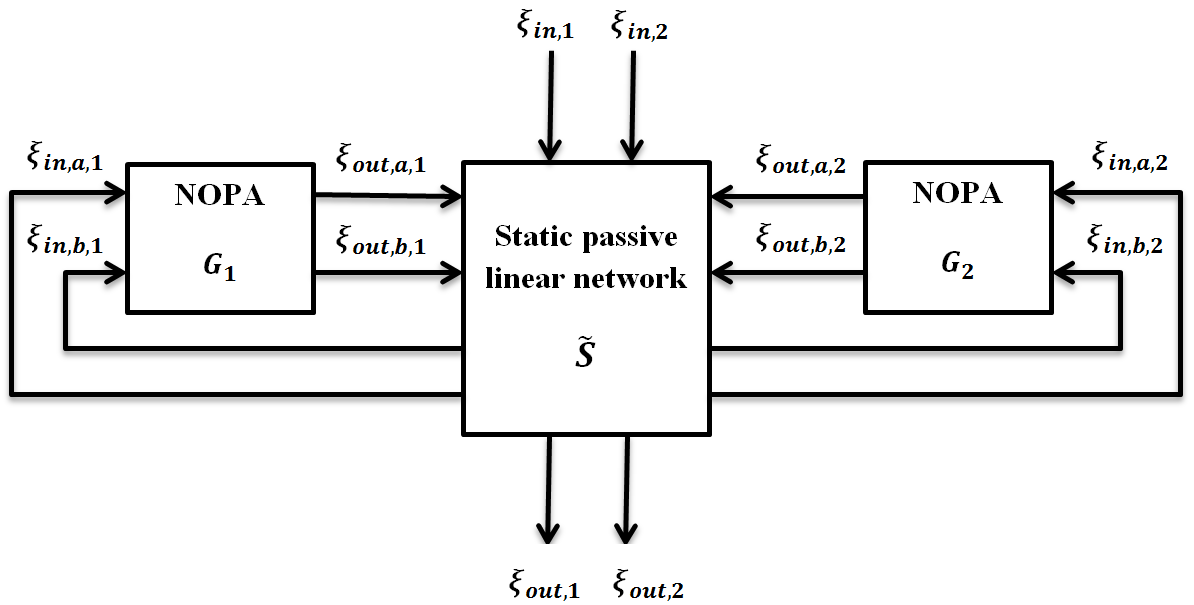}
\caption{System structure.}\label{fig:system}
\end{center}
\end{figure}
Coherent feedback is a feedback loop which directly connects quantum optical components without employing any measurement apparatus in the loop \cite{GW2009}. Our previous work \cite{SN2013} presents a dual-NOPA coherent feedback scheme comprised by two NOPAs as shown in Fig. \ref{fig:dual-NOPA-cfb}. 
Each NOPA $G_i~(i=1,2)$ in the figure is simply denoted by a block with four inputs and two outputs. $G_i$ contains two modes $a_i$ and $b_i$. Ingoing signal $\xi_{in,a,i}$ and amplification loss $\xi_{loss,a,i}$  interact with mode $a_i$; similarly $\xi_{in,b,i}$ and amplification loss $\xi_{loss,b,i}$ interact with mode $b_i$. The output $\xi_{out,a,1}$ of $G_1$ is connected to the input $\xi_{in,a,2}$ to $G_2$, and the outgoing signal $\xi_{out,b,2}$ from $G_2$ is the input to $G_1$. EPR entanglement is generated between outputs $\xi_{out,a,2}$ and $\xi_{out,b,1}$. 
A more detailed description of the NOPA will be given in Section \ref{sec:NOPA}. 
EPR entanglement generated from a dual-NOPA coherent feedback network where two NOPAs are placed at two endpoints (Alice and Bob) separately is compared to that of a single NOPA located in the middle (at Charlie's), at a location between Alice's and Bob's, in \cite{SN2013}. When amplification losses are neglected, under the same values of configuration parameters such as transmission losses, decay rates and pump power, the coherent feedback network improves EPR entanglement between two outgoing fields in terms of increasing the level of two-mode squeezing that can be achieved over the single NOPA, see \cite{SN2013}. Also, with the same setting of decay rates and when losses are neglected, the coherent feedback network requires less pump power to generate the same degree of two-mode squeezing compared to the single NOPA. Thus the coherent feedback scheme has improvement in EPR entanglement generation.

Based on the above facts, we consider a system consisting of two NOPAs connected in a coherent feedback loop with a static passive linear network which is denoted by a $6 \times 6$ complex unitary matrix $\tilde{S}$, as shown in Fig. \ref{fig:system}. The passive linear network can be assembled by several static linear optical devices, such as beamsplitters and phase shifters, of which the transformation functions are $2 \times 2$ unitary matrices \cite{NJD2009}. The EPR entanglement between the continuous-mode fields $\xi_{out,1}$ and $\xi_{out,2}$ is of interest. 
The coherent feedback network in Fig. \ref{fig:dual-NOPA-cfb} corresponds to a special case where $\tilde{S}$ takes on a particular value that will be given in Eq.~(\ref{eq:Scfb}) in Section \ref{sec:system-model}. 
EPR entanglement is quantified by the amount of two-mode squeezing between the two fields at the frequency $\omega=0$ rad/s. The two-mode squeezing will be given by a certain nonnegative-valued function $V(0; \tilde{S})$ of the matrix $\tilde{S}$ (to be defined in Section II-B), and strong EPR entanglement corresponds to a small value of this function.  
Thus, the aim of the paper is to optimize the two-mode squeezing by finding a local minimum (denoted by $\tilde{S}_{lm}$) of the real-valued function $V(0; \tilde{S})$ subject to the constraint that $\tilde{S}$ is unitary, which can be solved by a modified steepest descent algorithm on a Stiefel manifold as proposed in \cite{M2002}, with $\tilde{S}_{cfb}$ as an initial point. Via the decomposition of $\tilde{S}_{lm}$ into the product of  $15$ two-level unitary matrices \cite{LRY2013}, we can then find the configurations of optical devices that realizes the passive linear network $\tilde{S}_{lm}$.

The structure of the rest of this paper is as follows. A brief review about linear quantum systems, EPR entanglement between two continuous-mode fields and dynamics of a NOPA is given in Section \ref{sec:prelim}. Section \ref{sec:system-model} describes the system of interest and Section \ref{sec:optimization} explains the optimization process. In Section \ref{sec:decomposition }, by a decomposition of the unitary matrix $\tilde{S}_{lm}$, a detailed physical configuration of the whole network is presented. Finally Section \ref{sec:conclusion} gives a short conclusion of this paper.

\section{Preliminaries}
\label{sec:prelim}
The notations used in this paper are as follows: $\imath=\sqrt{-1}$ and $\operatorname{Re}$ denotes the real part of a complex number. The conjugate of a matrix is denoted by $\cdot^\#$, $\cdot^T$ denotes the transpose of a matrix of numbers or operators and $\cdot^*$ denotes (i) the complex conjugate of a number, (ii) the conjugate transpose of a matrix, as well as (iii) the adjoint of an operator. $O_{m\times n}$ is an $m$ by $n$ zero matrix and $I_n$ is an $n$ by $n$ identity matrix. Trace operator is written as $\operatorname{Tr[\cdot]}$ and tensor product is $\otimes$. Also, $\operatorname{eig}(\cdot)$ denotes eigenvalues of a matrix and $\max(\cdot)$ denotes the maximum value.

\subsection{Linear quantum systems}
\label{sec:linear-qsys}
An open linear quantum system without a scattering process contains $n$-bosonic modes $a_j(t)~(j=1,\ldots, n)$ satisfying $[a_i(t), a_j(t)^*]=\delta_{ij}$. The system interacts its environment via a time-varying interaction Hamitonian
\small
\begin{equation}
\label{interaction}
   H_{\rm int}(t) = \imath \sum_{j=1}^m (L_j\xi_j(t)^* - L_j^* \xi_j(t)), 
\end{equation}
\normalsize
where $L_j$ is the $j$-th system coupling operator and  $\xi_j(t)~(j=1,\ldots,m)$ is the field operator describing the $j$-th environment field  \cite{GardinerBook}. When the environment is under the condition of the Markov limit, the field operators satisfy $[\xi_j(t), \xi_j(s)^*]=\delta(t-s)$, where $\delta(t)$ denotes the Dirac delta function. 
When $L_j$ is linear and $H$ is quadratic in $a_j$ and $a_j^*$, the Heisenberg evolutions of mode $a_j$ and output filed operator $\xi_{out,j}$ are defined by $a_j(t)=U(t)^* a_j U(t)$  and $\xi_{out,j}(t)=U(t)^*\xi_{in,j}(t)U(t)$ with unitary  $U(t)={\rm exp}^{\hspace{-0.5cm}\longrightarrow}~
(-i\int_0^t H_{\rm int}(s)ds)$ and satisfy \cite{Belavkin2008}, \cite{WisemanBook}
\small
\begin{eqnarray}
    \dot{z}(t)&=&Az(t)+B\xi(t), \label{eq:dynamics} \\
     \xi_{out,j}(t)&=&Cz(t)+D\xi(t). \label{eq:output}
\end{eqnarray}
\normalsize
where
\small
\begin{eqnarray}
    z&=&(a_1^q, a_1^p, \ldots, a_n^q, a_n^p)^T, \nonumber\\
    \xi &=&(\xi_{1}^q, \xi_{1}^p, \ldots, \xi_{m}^q, \xi_{m}^p)^T, \nonumber\\
    \xi_{out}&=&(\xi_{out,1}^q, \xi_{out,1}^p, \ldots, \xi_{out,l}^q, \xi_{out,l}^p)^T,
\end{eqnarray}
\normalsize
with {\it quadratures} \cite{Belavkin2008,WisemanBook}
\small
\begin{eqnarray}
a_j^q &=& a_j+a_j^*, \quad a_j^p = (a_j-a_j^*)/i, \nonumber \\
\xi_j^q &=& \xi_j+\xi_j^*, \quad \xi_j^p = (\xi_j-\xi_j^*)/i. \label{eq: quadratures}
\end{eqnarray}
\normalsize

\subsection{EPR entanglement between two continuous-mode fields}
\label{sec:entanglement}
It is important to note that the output fields $\xi_{out,1}$ and $\xi_{out,2}$ are two continuous-mode Gaussian fields rather than two single mode Gaussian fields. That is, each of $\xi_{out,1}$ and $\xi_{out,2}$ contain a continuum of modes rather than just a single mode. Therefore, the entanglement of the fields cannot be measured using entanglement measures for bipartite Gaussian systems, such as the well-known logarithmic negativity measure \cite{Laurat2005}. Instead, when the incoming fields are in the vacuum state, the EPR entanglement of $\xi_{out,1}$ and $\xi_{out,2}$ is assessed in the frequency domain using two functions $V_+(\imath\omega)$  and $V_-(\imath\omega)$ \cite{BL2005,Ou1992,Vitali2006} that will be detailed below .

$F(\imath\omega)$, the Fourier transform of $f(t)$ is defined as $F\left(\imath\omega\right)=\frac{1}{\sqrt{2\pi}}\int_{-\infty}^{\infty} f\left(t\right)e^{-\imath\omega t} dt$. Similarly, we get the Fourier transforms of $\xi_{out,1}(t)$, $\xi_{out,2}(t)$, $z(t)$ and $\xi(t)$, 
as  $\tilde \Xi_{out,1}\left(\imath\omega\right)$, $\tilde \Xi_{out,2}\left(\imath\omega\right)$, $Z(\imath \omega)$  and  $\Xi(\imath \omega)$, respectively. 
Using (\ref{eq:dynamics}), (\ref{eq:output}) and the definition of the Fourier transform, we get
\small
\begin{eqnarray}
&\tilde \Xi_{out,1}^q(\imath \omega)+\tilde \Xi_{out,2}^q(\imath \omega)=\int_{-\infty}^{\infty} \xi_{out,1}^q(t)e^{-\imath\omega t} dt+\int_{-\infty}^{\infty} \xi_{out,2}^q(t)e^{-\imath\omega t}  dt =C_1 Z\left(\imath\omega\right)+ D_1\Xi\left(\imath\omega\right), \nonumber\\
&\tilde \Xi_{out,1}^p(\imath \omega)-\tilde \Xi_{out,2}^p(\imath \omega)= \int_{-\infty}^{\infty} \xi_{out,1}^p(t)e^{-\imath\omega t} dt-\int_{-\infty}^{\infty} \xi_{out,2}^p(t)e^{-\imath\omega t}  dt =C_2 Z\left(\imath\omega\right)+ D_2\Xi\left(\imath\omega\right),
\end{eqnarray}
\normalsize
where $ C_1=[1\ 0\ 1\ 0]C$, $C_2=[0\ 1\ 0\ {-}1]C$, $D_1=[1\ 0\ 1\ 0]D$ and $D_2=[0\ 1\ 0\ {-}1]D$.

The two-mode squeezing spectra $V_{+}(\imath \omega)$ and $V_{-}(\imath \omega)$ are defined as
\small
\begin{align}
& \langle (\tilde \Xi_{out,1}^q(\imath \omega)+\tilde \Xi_{out,2}^q(\imath \omega))^* (\tilde \Xi_{out,1}^q(\imath \omega')+\tilde \Xi_{out,2}^q(\imath \omega')) \rangle = V_+(\imath \omega)\delta(\omega-\omega'), \nonumber \\
& \langle (\tilde \Xi_{out,1}^p(\imath \omega)-\tilde \Xi_{out,2}^p(\imath \omega))^* (\tilde \Xi_{out,1}^p(\imath \omega')-\tilde \Xi_{out,2}^p(\imath \omega')) \rangle  =  V_-(\imath \omega) \delta(\omega-\omega'),
\end{align}
\normalsize
where $\langle \cdot \rangle$ denotes quantum expectation. 
$V_+(\imath \omega)$ and $V_-(\imath \omega)$ are real valued and can be easily calculated as described in \cite{NY2012,GJN2010},
\small
\begin{align}
V_+(\imath\omega)=& {\rm Tr}\left[H_1(\imath\omega)^* H_1(\imath\omega)\right], \label{eq:V_+}\\
V_-(\imath\omega)=& {\rm Tr}\left[H_2(\imath\omega)^* H_2(\imath\omega)\right], \label{eq:V_-}
\end{align}
\normalsize
where $H_1$ and $H_2$ are transfer functions
\small
\begin{align}
H_j(\imath\omega)=C_{j}\left(\imath\omega I-A \right)^{-1}B+D_{j},  ~~(j=1,2). \label{eq:transfer-function}
\end{align}
\normalsize
Denote $V(\imath\omega) = V_+(\imath\omega)+V_-(\imath\omega)$. A sufficient condition for the fields $\xi_{out,1}$ and  $\xi_{out,2}$ to be EPR-entangled at the frequency $\omega$ rad/s is \cite{Vitali2006},
\small
\begin{align}
V(\imath\omega)< 4. \label{eq:entanglement-criterion}
\end{align}
\normalsize
Ideally, we would like $V(\imath\omega) = V_{\pm}(\imath\omega)= 0$ for all $\omega$, which denotes infinite-bandwidth two-mode squeezing, representing an ideal Einstein-Podolski-Rosen state. However, in reality the ideal EPR correlation can not be achieved,  so in practice the goal is to make $V(\imath \omega)$ as small as possible over a wide frequency range \cite{Vitali2006}. 

Define $\xi^{\psi_1}_{out,1}=e^{\imath \psi_1}\xi_{out,1}$, $\xi^{\psi_2}_{out,2}=e^{\imath \psi_2}\xi_{out,2}$ with $\psi_1, \psi_2 \in(-\pi,\pi]$ and denote the corresponding two-mode squeezing spectra as $V^{\psi_1, \psi_2}_\pm(\imath\omega,\psi_1, \psi_2)$, we have the following definition of EPR entanglement.
\begin{definition}
Fields $\xi_{out,1}$ and  $\xi_{out,2}$ are EPR entangled at the frequency $\omega$ rad/s if $\exists ~\psi_1, \psi_2 \in(-\pi,\pi]$ such that
\small
\begin{align}
 V^{\psi_1, \psi_2}_+(\imath\omega,\psi_1, \psi_2)+V^{\psi_1, \psi_2}_-(\imath\omega,\psi_1, \psi_2)< 4. \label{eq:entanglement-criterion-2}
\end{align}
\normalsize
Unless otherwise specified, throughout the paper, EPR entanglement refers to the case with $\psi_1=\psi_2=0$.
EPR entanglement is said to vanish at $\omega$ if there are no values of $\psi_1$ and $\psi_2$ satisfying the above criterion.
\end{definition}

Following \cite{SN2013, NY2012}, we have a good approximation $V_{+}(i\omega) \approx V_{+}(0)$ and $V_{-}(i\omega) \approx V_{-}(0)$ at low frequencies. Thus in the rest of paper, we focus on $V_{\pm}(0)$ at frequency $\omega=0$ as a measure to quantify EPR entanglement.  

\subsection{The nondegenerate optical parametric amplifier (NOPA) and the dual-NOPA coherent-feedback network}
\label{sec:NOPA}
\begin{figure}[htbp]
\begin{center}
\includegraphics[scale=0.4]{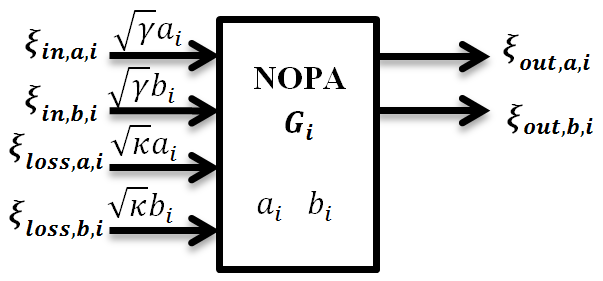}
\caption{Input/output block representation of a NOPA.}\label{fig:single-NOPA}
\end{center}
\end{figure}

A NOPA ($G_i$) is a linear quantum system with four ingoing fields in the vacuum state and two outgoing fields, as shown in Fig. \ref{fig:single-NOPA}.
By shining the pump beam onto the $\chi^{(2)}$ nonlinear crystal, the two oscillator modes $a_i$ and $b_i$ inside the cavity become coupled via the two-mode squeezing Hamiltonian $H= \frac{\imath}{2} \epsilon\left( a_i^* b_i^*- a_ib_i\right)$, where $\epsilon$ is a real coupling coefficient related to the amplitude of the pump beam \cite{Ou1992}.
The modes satisfy the commutation relations $[a_i, a_i^*]=1$, $[b_i, b_i^*]=1$, $[a_i, b_i]=0$, and $[a_i, b_i^*]=0$ \cite{GardinerBook}. Mode $a_i$ is coupled to ingoing noise $\xi_{in,a,i}$ and amplification loss $\xi_{loss,a,i}$ via the coupling operators  $L_1=\sqrt{\gamma}a$ and $L_3=\sqrt{\kappa}a$, respectively, for some constant damping rates $\gamma$ and $\kappa$; similarly mode $b_i$  interacts with input signal $\xi_{in,b,i}$ and additional noise $\xi_{loss,b,i}$ by operators $L_2=\sqrt{\gamma}b$ and $L_4=\sqrt{\kappa}b$. The dynamics of the NOPA ($G_i$) is
\small
\begin{align}
\dot{a_i}\left(t\right)=&-\left(\frac{\gamma+\kappa}{2}\right)a_i\left(t\right)+\frac{\epsilon}{2}b_i^*\left(t\right)-\sqrt{\gamma}\xi_{in,a,i}\left(t\right) -\sqrt{\kappa}\xi_{loss,a,i}\left(t\right),\nonumber \\
\dot{b_i}\left(t\right)=&-\left(\frac{\gamma+\kappa}{2}\right)b_i\left(t\right)+\frac{\epsilon}{2}a_i^*\left(t\right)-\sqrt{\gamma}\xi_{in,b,i}\left(t\right)-\sqrt{\kappa}\xi_{loss,b,i}\left(t\right), \label{eq:NOPA-dynamics1}
\end{align}
\normalsize
with outputs
\small
\begin{align}
\xi_{out,a,i}\left(t\right)=&\sqrt{\gamma}a_i\left(t\right)+\xi_{in,a,i}\left(t\right),\nonumber \\
\xi_{out,b,i}\left(t\right)=&\sqrt{\gamma}b_i\left(t\right)+\xi_{in,b,i}\left(t\right). \label{eq:NOPA-dynamics2}
\end{align}
\normalsize

More details of the standard NOPA set up can be found in \cite{Ou1992}.
\section{The system model}
\label{sec:system-model}
We consider the system shown in Fig. \ref{fig:system}. The whole network consists of two NOPAs and a static passive linear sub-system denoted by $\tilde S$. The sub-system has six input signals $\xi_{in,1}$, $\xi_{in,2}$, $\xi_{out,a,1}$, $\xi_{out,b,1}$, $\xi_{out,a,2}$ and $\xi_{out, b, 2}$, among which $\xi_{out,a,1}$, $\xi_{out,b,1}$ are outputs of NOPA1, and $\xi_{out,a,2}$, $\xi_{out, b, 2}$ are outputs of NOPA2. Among the six outgoing beams of the passive sub-system are the two ingoing signals into NOPA1 $\xi_{in,a,1}$ and $\xi_{in,b,1}$, the two input beams of NOPA2 $\xi_{in,a,2}$ and $\xi_{in,b,2}$, and the EPR entanglement of interest is between $\xi_{out,1}$ and $\xi_{out,2}$. The ingoing fields of the passive linear network are all in the vacuum state \cite{NJD2009}. 
The transfer function of the static sub-system is a $6 \times 6$  matrix $\tilde S$, thus we have
\small
\begin{eqnarray}
\left[ \begin{array}{c}
\xi_{out,1}\\ \xi_{out,2}\\ \xi_{in,a,1}\\ \xi_{in,b,1}\\ \xi_{in,a,2}\\ \xi_{in,b,2} \end{array} \right] 
= \tilde S \left[ \begin{array}{c}
\xi_{in,1}\\ \xi_{in,2}\\ \xi_{out,a,1}\\ \xi_{out,b,1}\\ \xi_{out,a,2}\\ \xi_{out,b,2} \end{array} \right] \label{eq:S}
\end{eqnarray}，
\normalsize
where $\tilde S$ is a complex unitary matrix \cite{NJD2009},
\small
\begin{equation}
\tilde S^* \tilde S = \tilde S \tilde S^*= I_6. \label{eq： complex-unitary}
\end{equation} 
\normalsize

For the static passive linear matrix for the dual-NOPA coherent feedback network \cite{SN2013} shown in Fig. \ref{fig:dual-NOPA-cfb}, the matrix $\tilde S$ is
\small
\begin{eqnarray}
\tilde{S}_{cfb} &=& \left[\begin{array}{cccccc} 0 & 0 & 0 & 0 & 1 & 0\\ 0 & 0 & 0 & 1 & 0 & 0\\1 & 0 & 0 & 0 & 0& 0 \\0 & 0 & 0 & 0& 0 & 1 \\ 0 & 0 & 1 & 0& 0 & 0 \\ 0 & 1 & 0 & 0& 0 & 0 \\ \end{array}\right].
\label{eq:Scfb} \end{eqnarray}
\normalsize

Both NOPAs ($G_1$ and $G_2$) in the network are identical with the same coupling constants $\epsilon$, $\gamma$ and $\kappa$  as discussed in Section \ref{sec:NOPA}. $G_1$ has two modes $a_1$ and $b_1$, and $G_2$ contains modes $a_2$ and $b_2$. The oscillation modes follow the commutation relations $[a_i, a_j^*]=\delta_{ij}$, $[a_i, b_j]=0$,  $[a_i, b_j^*]=0$,  $[a_i, a_j]=0$ and $[b_i, b_j]=0$ $(i,j=1,2)$.

Define the following quadratures
\small
\begin{align}
z=&[a^q_1, a^p_1, b^q_1, b^p_1,a^q_2, a^p_2, b^q_2, b^p_2]^T,\nonumber \\
\xi_{loss}=&[\xi^q_{loss,a,1},\xi^p_{loss,a,1},\xi^q_{loss,b,1},\xi^p_{loss,b,1},\xi^q_{loss,a,2},\xi^p_{loss,a,2},\xi^q_{loss,b,2},\xi^p_{loss,b,2}]^T, \nonumber\\
\xi^{(i)}=&[\xi^q_{in,1},\xi^p_{in,1},\xi^q_{in,2},\xi^p_{in,2}]^T, \nonumber\\
\xi^{(o)}=&[\xi^q_{out,1},\xi^p_{out,1},\xi^q_{out,2},\xi^p_{out,2}]^T, \nonumber\\
\xi =&[{\xi^{(i)}}^T, \xi_{loss}^T]^T.
\end{align}
\normalsize
Define the real unitary matrix $S$ as the quadrature form of matrix $\tilde S$. Based on the defining equations for the quadratures (\ref{eq: quadratures}), the relation between $S$ and $\tilde{S}$ is 
\small
\begin{align}
S=\frac{1}{2}K\tilde{S}K^* + \frac{1}{2}K^\# \tilde{S}^\# K^T,  \label{eq:relations-real-complex-matrix}
\end{align}
\normalsize
where 
\small
\begin{eqnarray}
K= I_6 \otimes \left[ \begin{array}{c} 1\\ -\imath  \end{array} \right]. \label{eq:K}
\end{eqnarray}
\normalsize

According to the dynamics of the two NOPAs given by (\ref{eq:NOPA-dynamics1}) and (\ref{eq:NOPA-dynamics2}), and similar to the discussion in Section \ref{sec:linear-qsys}, we have
\small
\begin{align}
\dot{z}\left(t\right)=&A z\left(t\right)+B\xi\left(t\right), \nonumber \\
\xi^{(o)}\left(t\right)=&C z\left(t\right)+D\xi\left(t\right).\label{eq: system-dynamics}
\end{align}
\normalsize
$A$, $B$, $C$ and $D$ are real matrices
\small
\begin{eqnarray}
A &=& R-\gamma (X-I_8)  \nonumber \\
B &=& \left[\begin{array}{cc} -\sqrt{\gamma}XS_{21} &  -\sqrt{\kappa}I_8 \end{array}\right]  \nonumber \\
C &=& \sqrt{\gamma}S_{12}X  \nonumber \\
D &=& \left(S_{11} +S_{12} X S_{21} \right)\left[\begin{array}{cc} I_4 &  O_{4 \times 8}  \end{array}\right]  \label{eq:abcd}
\end{eqnarray} 
\normalsize
where\\
\small
\begin{eqnarray}
R&=&\left[\begin{array}{cccccccc}
-\frac{\gamma+\kappa}{2} & 0 & \frac{\epsilon}{2} & 0 & 0 & 0 & 0 & 0 \\
0 & -\frac{\gamma+\kappa}{2} & 0 & -\frac{\epsilon}{2} & 0 & 0 & 0 & 0 \\
\frac{\epsilon}{2} & 0 & -\frac{\gamma+\kappa}{2} & 0 & 0 & 0 & 0 & 0 \\
0 & -\frac{\epsilon}{2} & 0 & -\frac{\gamma+\kappa}{2} & 0 & 0 & 0 & 0 \\
0 & 0 & 0 & 0 & -\frac{\gamma+\kappa}{2} & 0 & \frac{\epsilon}{2} & 0 \\
0 & 0 & 0 & 0 & 0 & -\frac{\gamma+\kappa}{2} & 0 & -\frac{\epsilon}{2} \\
0 & 0 & 0 & 0 & \frac{\epsilon}{2} & 0 & -\frac{\gamma+\kappa}{2} & 0  \\
0 & 0 & 0 & 0 & 0 & -\frac{\epsilon}{2} & 0 & -\frac{\gamma+\kappa}{2} 
\end{array} \right],\nonumber \\
X &=& \left(I_8-S_{22}\right)^{-1}, \nonumber \\
S_{11} &=& \left[\begin{array}{cc} I_4 &  O_{4 \times 8} \end{array}\right] S \left[\begin{array}{c} I_4 \\  O_{8 \times 4} \end{array}\right], \nonumber \\
S_{21} &=& \left[\begin{array}{cc} O_{8 \times 4} &  I_8 \end{array}\right] S \left[\begin{array}{c} I_4 \\  O_{8 \times 4} \end{array}\right], \nonumber\\
S_{12} &=& \left[\begin{array}{cc} I_4 &  O_{4 \times 8} \end{array}\right] S \left[\begin{array}{c} O_{4 \times 8} \\ I_8  \end{array}\right],\nonumber \\
S_{22} &=& \left[\begin{array}{cc} O_{8 \times 4} & I_8  \end{array}\right] S \left[\begin{array}{c} O_{4 \times 8} \\ I_8 \end{array}\right] \label{eq:matrices}. 
\end{eqnarray}
\normalsize
More details of how to obtain (\ref{eq: system-dynamics}) are given in Appendix~1.
Based on (\ref{eq:V_+}), (\ref{eq:V_-}) and (\ref{eq:transfer-function}), we have
\small
\begin{eqnarray}
H &=& D-CA^{-1}B\nonumber \\
H_1 &=& \left[\begin{array}{cccc} 1 & 0 & 1 & 0 \end{array}\right]H, \nonumber \\
H_2 &=& \left[\begin{array}{cccc} 0 & 1 & 0 & -1 \end{array}\right]H, \label{eq:H}
\end{eqnarray} 
\normalsize
and the two-mode squeezing spectra are
\small
\begin{eqnarray}
V(0) &=& V_+(0)+V_-(0)=\operatorname{Tr}\left[H_1^*H_1+H_2^*H_2\right]=\operatorname{Tr}\left[H^*M_{1,2} H  \right], \label{eq:entanglement}
\end{eqnarray}
\normalsize
where
\small
\begin{eqnarray}
M_{1,2} &=&  \left[\begin{array}{cccc} 1 & 0 & 1 & 0 \\ 0 & 1 & 0 & -1 \\1 & 0 & 1 & 0 \\0 & -1 & 0 & 1 \end{array}\right].  
\end{eqnarray}
\normalsize

Since $V(0)$ is parametrized by the matrix $\tilde S$ or $S$, we define $V(0; \tilde S)$ as the value of $V(0)$ for a fixed value of $\tilde S$, and $V(0; S)$ as the value of $V(0)$ for a fixed value of $S$. 
\section{Optimization of $\tilde S$}
\label{sec:optimization}
Using the same parameters (total pump power and damping rates) as that of the coherent-feedback dual NOPA system described in \cite{SN2013}, for each NOPA we set the constant relating to the amplitude of the pump beam $\epsilon = 0.4 \gamma_r$  and the damping rate $\gamma= \gamma_r$, where  $\gamma_{r}=7.2\times10^7$ Hz is the reference value for the transmissivity rate of the mirrors. We consider the system in the ideal case, where there are no losses ($\kappa=0$). Following  Section \ref{sec:system-model}, we compute the two-mode squeezing of the dual-NOPA coherent feedback network to be $10\log_{10}  V(0; \tilde{S}_{cfb}) = -26.235$ dB.

In this section, we aim to find a complex unitary matrix $\tilde{S}_{lm}$ at which the cost function $V(0; \tilde{S})$ is locally minimized. We will numerically solve this optimization problem using the method of modified steepest descent on a Stiefel manifold, which reformulates the problem with a unitary constraint as an unconstrained problem on a Stiefel manifold. The Stiefel manifold in our problem is the set $ St(6,6)=\left\lbrace\tilde{S} \in \mathbb{C}^{6 \times 6} : \tilde{S}^* \tilde{S} = I\right\rbrace$.  The modified steepest descent on a Stiefel manifold method employs the first-order derivative of the cost function, more details about this algorithm can be found in \cite{M2002} . 

For any square matrix $Y$ such that $I-Y$ is invertible, we have $(I-Y)^{-1}=(I-Y)^{-1}(I+Y-Y)=I+(I-Y)^{-1}Y$. Based on the above fact and equations (\ref{eq:sub-S}), (\ref{eq:abcd}) and (\ref{eq:H}), we expand $H(S +\Delta S)$ as $H(S)+ H(\Delta S)+ O(\Delta S^2)$, where $O(\Delta S^2)$ denotes terms that depend on terms that are products containing at least two $\Delta S$.  

By using (\ref{eq:entanglement}), we have
\small
\begin{eqnarray}
\quad V(0; S +\Delta S)&=&\operatorname{Tr} \left[H(S+\Delta S)^* M_{1,2} H(S+\Delta S) \right] \nonumber\\
&=& V(0; S) + \operatorname{Tr}[H(\Delta S)^*M_{1,2} H(S) + H(S)^*M_{1,2} H(\Delta S) ] + O(\lVert\Delta S\rVert^2), 
\end{eqnarray}
\normalsize
where  $O(\lVert\Delta S\rVert^2)$ denotes that the function $O(\lVert\Delta S\rVert^2)$ satisfies $\frac{O(\lVert\Delta S\rVert^2)}{\lVert\Delta S\rVert^2} \leq c$ for some positive constant $c$ for all $\lVert\Delta S\rVert>0$ sufficiently small. Since $H(\Delta S)$ and $H(S)$ are real matrices at $\omega=0$, $M_{1,2}$ is a real symmetric matrix, and a matrix and its transpose have the same trace, we get
\small
\begin{eqnarray}
V(0; S+\Delta S)= V(0; S) + 2\operatorname{Tr}[M H(\Delta S)] +O(\lVert\Delta S\rVert^2),
\end{eqnarray}
\normalsize
where $M=H(S)^*M_{1,2}$. Based on  (\ref{eq:relations-real-complex-matrix}) and the property that trace is invariant under cyclic permutations, we have
\small
\begin{eqnarray}
V(0; \tilde{S} + \Delta \tilde{S})=V(0; \tilde{S})+ \operatorname{Re}\operatorname{Tr}[\Delta\tilde{S}^*D_{\tilde{S}}] + O(\lVert\Delta\tilde S\rVert^2),
\end{eqnarray}
\normalsize
where 
\small
\begin{eqnarray}
D_{\tilde{S}} &=& 2K^* N^* K , \label{eq: Ds}\\
N &=& \left(\left[\begin{array}{c} I_4 \\ O_{8 \times 4} \end{array}\right]\left[\begin{array}{cc} I_4 & O_{4 \times 8} \end{array}\right] M +\left[\begin{array}{c}O_{4 \times 8} \\ I_8 \end{array}\right]X S_{21}\left[\begin{array}{cc} I_4 & O_{4 \times 8} \end{array}\right] M  \right. \nonumber \\
&& \quad \left. -\sqrt{\gamma}\left[\begin{array}{c}O_{4 \times 8} \\ I_8 \end{array}\right]X A^{-1}B M \right)  \left( \left[\begin{array}{cc} I_4 & O_{4 \times 8} \end{array}\right] + S_{12} X \left[\begin{array}{cc}  O_{8 \times 4} & I_8 \end{array}\right] \right. \nonumber \\
&& \quad  \left. +\sqrt{\gamma}C A^{-1} X  \left[\begin{array}{cc}  O_{8 \times 4} & I_8 \end{array}\right] \right) \label{eq: N}.
\end{eqnarray}
\normalsize
When $\Delta \tilde{S}$ approaches the zero matrix $O_{6,6}$, $D_{\tilde{S}}$ is the directional derivative of $V(0; \tilde{S})$ at $\tilde{S}$ in the direction $\Delta \tilde S$ \cite{M2002}.

\vspace{6pt}
By applying the modified steepest descent on Stiefel manifold method, we use the following steps to find a matrix $\tilde{S}_{lm}$ at which the dual-NOPA system is stable and $V(0; \tilde S)$ is locally minimized. The system is stable if and only if the matrix $A$ in equation (\ref{eq: system-dynamics}) is Hurwitz, that is, real parts of all the eigenvalues of $A$ are negative. Denote the vector of all the mode operators of our system by $\tilde{z}(t)=[a_1(t), b_1(t), a_2(t), b_2(t)]^T$ and define the intra-cavity photon number operator as $n(t)=\tilde{z}(t)^*\tilde{z}(t)$. If the system is stable, we have $\lim_{t\to\infty}\langle \tilde{z}(t) \rangle=0$ \cite{WisemanBook}. Moreover, based on the quantum Ito rules \cite{Hudson1984}, Section~2.5 in \cite{Crisafulli_thesis} verifies that $\lim_{t\to\infty}\langle n(t)\rangle$ is bounded. Therefore, a stable system physically means that the mean total number of intra-cavity photons in the system would not keep increasing as time $t$ approaches infinity. Note that, based on (\ref{eq:abcd}) and (\ref{eq:H}), to get $V(0; \tilde S)$, $(I_8-S_{22})$ and $A$ must be invertible. The algorithm is initiated at the point $\tilde{S}_{cfb}$ given by (\ref{eq:Scfb}), the static passive linear matrix denoting the dual-NOPA coherent feedback network shown in Fig. \ref{fig:dual-NOPA-cfb}.

{\it Step 1}. Start from $\tilde{S}= \tilde{S}_{cfb}$, choose step size $\rho=1$. 

{\it Step 2}. Calculate $D_{\tilde{S}}$ , the directional derivative of  $V(0; \tilde{S})$, by using equation (\ref{eq: Ds}), and compute the descent direction $Z=\tilde{S}D_{\tilde{S} }^* \tilde{S} - D_{\tilde{S} }$.

{\it Step 3}. Calculate $ \langle Z,Z \rangle  = {\rm Tr}[Z^*(I-\frac{1}{2}\tilde{S}\tilde{S}^*)Z]$. If $\sqrt{ \langle Z,Z \rangle }$ is smaller than $10^{-3}$, stop and set $\tilde{S}_{lm}= \tilde{S}$.

{\it Step 4}. Calculate $\tilde{S}_1=\pi (\tilde{S} +2\rho Z)$ (Note: if the singular value decomposition (SVD) of a matrix $X$ is $X=U \Sigma V^*$, then $\pi (X) =UIV^*$).  Calculate $A$ and $(I_8-S_{22})$ corresponding to $\tilde{S}_1$ based on (\ref{eq:abcd}) and (\ref{eq:H}).
If $\max(\operatorname{Re}({\rm eig}(A)))>0$ or $\det(A)=0$ or $\det(I_8-S_{22})=0$， go to step 5. Otherwise, if $V(\tilde{S})-V(\tilde{S}_1) \geq \rho \langle Z,Z \rangle $, set $2\rho \rightarrow \rho$, and go back to Step 4.

{\it Step 5}. Calculate $\tilde{S}_2=\pi (\tilde{S} +\rho Z)$. Compute $A$ and $(I_8-S_{22})$ corresponding to $\tilde{S}_2$. While $\max(\operatorname{Re}({\rm eig}(A)))>0$ or $\det(A)=0$ or $\det(I_8-S_{22})=0$, set $\frac{1}{2} \rho \rightarrow \rho $, go back to step 5. If $ V(\tilde{S})-V(\tilde{S}_2) < \frac{1}{2} \rho \langle Z,Z\rangle $ , set $\frac{1}{2} \rho \rightarrow \rho $, and go back to Step 5. 

{\it Step 6}. Set $\tilde{S}=\tilde{S}_2$ and repeat Step 2.

Thus, we get $\tilde{S}_{lm}$ and the corresponding gradient $G=D_{\tilde{S}}$ shown as (\ref{eq:S-optimized}) and (\ref{eq:G-optimized}) in Appendix~2, the operator norm of $G$ is $ \parallel G  \parallel= 6.497\times 10^{-4}$, the norm of a tangent direction $Z$ is $\sqrt{ \langle Z,Z \rangle }=9.112 \times 10^{-4}$ and EPR entanglement is locally maximized for a locally minimum value of $V(0; \tilde{S})$. The locally minimized two-mode squeezing is $10\log_{10}V(\tilde{S}_{lm})= 10\log_{10}(4.1824\times 10^{-8})=-73.786$ dB, which is about $47.551$ dB less than the value reported in \cite{SN2013} for the dual NOPA coherent feedback system in Fig. \ref{fig:dual-NOPA-cfb} for the same values of the parameters of the NOPAs. Also, it is checked by equations (\ref{eq:V_+}), (\ref{eq:V_-}) and (\ref{eq:H}) that, $V_+(\tilde{S}_{lm}) \approx V_-(\tilde{S}_{lm})$. 

\section{Decomposition of unitary matrices}
\label{sec:decomposition }
As introduced in Section \ref{sec:intro}, the $6 \times 6$ matrix $ \tilde{S}_{lm}$ denotes a network constructed from  static passive linear optical devices. To find the specific physical configuration of the network, we first employ the approach in \cite{LRY2013} to decompose $\tilde{S}_{lm}$ as the product of $15$ two-level $6 \times 6$ unitary matrices, that is,  $ \tilde{S}_{lm}=\Pi_{k=1}^{15} \tilde S_k $. 
Here, a two-level $n \times n$ unitary matrix refers to a special type of unitary matrix that has a unitary $2 \times 2$ principal submatrix and the remaining matrix elements are the same as those of the $n \times n$ identity matrix. Each of these two-level unitary matrices has determinant with modulus $1$. The reason we use this method here is that any two-level unitary matrix is isomorphic to the set of $2 \times 2$ unitary matrices, which represent the transformation performed in the Heisenberg picture by static linear optical devices, such as beam splitters and phase shifters. 

The decomposition of $ \tilde{S}_{lm}$ does not give a unique group of two-level unitary matrices. Types of two-level unitary matrices in a group are determined by a vector $P=(p_1, p_2, \cdots, p_6)$, where entries correspond to a permutation of $(1, 2, \cdots, 6)$. A two-level matrix is named P-unitary matrix of type k if row and column indexes of its principal submatrix are $p_k$ and $p_{k+1}$. 
By setting $P=(6, 5 ,4, 3, 2, 1)$ and using the Matlab program {\it pub.m} developed by \cite{LRY2013}, we find a group of two-level unitary matrices $\tilde{S}_{k}$ ($k \in [1,13]$) shown as (\ref{eq:group_S_i}) in Appendix~2.

\begin{figure*}[htbp]
\begin{center}
\includegraphics[scale=0.45]{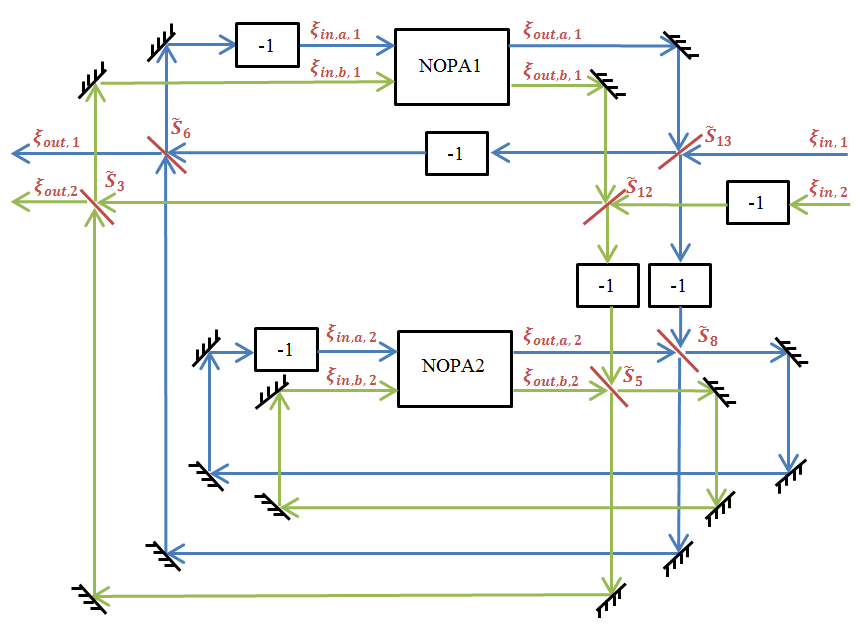}
\caption{Physical configuration of the optimized dual-NOPA network. The system contains NOPA1, NOPA2 and a static passive linear network which consists of six beam splitters $\tilde{S}_{k},k \in \lbrace 3, 5, 6, 8, 12, 13\rbrace$ (denoted by the red lines) with the parameters $\alpha_k$ as shown in equation (\ref{eq:alpha-beta}). For each NOPA, $\kappa=0$, $\epsilon = 0.4 \gamma_r$, $\gamma= \gamma_r$,  where  $\gamma_{\rm r}=7.2\times10^7$. The black lines are mirrors that are fully reflecting. Inputs fields $\xi_{in,1}$ and $\xi_{in,2}$ are in the vacuum state and EPR entanglement is generated between the two outputs $\xi_{out,1}$ and $\xi_{out,2}$ as discussed in Section \ref{sec:optimization}. }\label{fig:network-configuration}
\end{center}
\end{figure*}

A unitary matrix representing a beamsplitter has the form \cite{GK2005}
\begin{eqnarray}
\left[\begin{array}{cc}
\alpha & \beta \\
-\beta & \alpha
\end{array}\right],
\end{eqnarray}
where transmission rate $\alpha$ and reflection rate $\beta$ are real numbers satisfying $ \lvert \alpha \rvert ^2 + \lvert \beta \rvert ^2 =1$. Thus, $\tilde{S}_{k}$ ($k \in \lbrace 3, 5, 6, 8, 12, 13\rbrace$) represents transformation by a beamsplitter, 
with parameters
\begin{eqnarray}       
\alpha_3 &=& -\alpha_{13}= 0.9999632197,  \nonumber    \\  
\alpha_8 &=& -\alpha_5=0.0084711563, \nonumber  \\
\alpha_{12} &=& -\alpha_6 = 0.0123787627 , \nonumber  \\
\beta_k &=& \sqrt{1-\alpha_{k}^2}, ~~ k \in \lbrace 3, 5, 6, 8, 12, 13\rbrace .\label{eq:alpha-beta}
\end{eqnarray}   
The configuration of the whole network is shown in Fig. \ref{fig:network-configuration}. The network requires high accuracy of the value of the parameter $\alpha_k$. To achieve $10\log_{10}V(\tilde{S}_{lm})=-73.786$ dB, we need to keep at least six decimal places for $\alpha_k$. However, with a lower accuracy of less than six digits but more than one digit, we still get better two-mode squeezing than that of the dual-NOPA coherent feedback network. For example, by rounding off to two decimal places, that is, $\alpha_3= -\alpha_{13}= 1$, $\alpha_8= \alpha_{12} = -\alpha_5= -\alpha_6=0.01$, with $\beta_k = \sqrt{1-\alpha_{k}^2}$ ($k \in \lbrace 3, 5, 6, 8, 12, 13\rbrace$) as before, the two mode squeezing of the optimized network is $-36.546$ dB. When accuracy is less than two decimal digits, the network becomes exactly the dual-NOPA coherent feedback network.
\section{Conclusion}
\label{sec:conclusion}
This paper has studied the optimization of EPR entanglement in terms of maximising the two-mode squeezing generated by a quantum network that contains two NOPAs connected by a static passive linear network. The transformation implemented by the passive network is a  $6 \times 6$ complex unitary matrix. By employing the modified steepest descent on Stiefel manifold method, we have found the passive network $\tilde{S}_{lm}$ at which the two-mode squeezing function $V(0; \tilde{S})$ used to evaluate the EPR entanglement is approximately minimized locally. It is shown that $V_+(0; \tilde{S}_{lm}) \approx V_-(0; \tilde{S}_{lm})$ and $V(0; \tilde{S}_{lm})= 4.1824 \times 10^{-8}$, which approximates the ideal case of infinite squeezing where $V_+(0; \tilde{S})=V_-(0; \tilde{S})=0$. 
Also, with the same values of the parameters $\epsilon$ and $\gamma$ and without considering losses ($\kappa=0$ ), the optimized network improves the EPR entanglement by a significant reduction of $47.551$ dB in the two-mode squeezing compared to the one of the dual-NOPA coherent feedback network of Fig.~\ref{fig:dual-NOPA-cfb} studied in \cite{SN2013}. Finally, by decomposing $\tilde{S}_{lm}$ into a product of two-level unitary matrices, we have found the physical set up of the optimized network as shown in Fig. \ref{fig:network-configuration}.  The network requires that beamsplitters have highly accurate realization of the transmission rates $\alpha_k$, $k \in \lbrace 3, 5, 6, 8, 12, 13\rbrace$ with an accuracy of at least six decimal places to achieve the two-mode squeezing $10\log_{10} V(0; \tilde{S}_{lm})=-73.786$ dB. If accuracy is less than six and more than one decimal digits, the two-mode squeezing is lower than $-73.786$ dB but better than that of the dual-NOPA coherent feedback network in Fig. \ref{fig:dual-NOPA-cfb}. If accuracy of $\alpha_k$ is zero or one decimal digit, the network coincides with  the dual-NOPA coherent feedback network in Fig.~\ref{fig:dual-NOPA-cfb}.

\section*{APPENDIX 1}
For convenience, define the following quadratures
\small
\begin{align}
z=&[a^q_1, a^p_1, b^q_1, b^p_1,a^q_2, a^p_2, b^q_2, b^p_2]^T,\nonumber \\
\xi_{in}=&[\xi^q_{in,a,1},\xi^p_{in,a,1},\xi^q_{in,b,1},\xi^p_{in,b,1},\xi^q_{in,a,2},\xi^p_{in,a,2},\xi^q_{in,b,2},\xi^p_{in,b,2}]^T, \nonumber\\
\xi_{out}=&[\xi^q_{out,a,1},\xi^p_{out,a,1},\xi^q_{out,b,1},\xi^p_{out,b,1},\xi^q_{out,a,2},\xi^p_{out,a,2},\xi^q_{out,b,2},\xi^p_{out,b,2}]^T, \nonumber\\
\xi_{loss}=&[\xi^q_{loss,a,1},\xi^p_{loss,a,1},\xi^q_{loss,b,1},\xi^p_{loss,b,1},\xi^q_{loss,a,2},\xi^p_{loss,a,2},\xi^q_{loss,b,2},\xi^p_{loss,b,2}]^T, \nonumber\\
\xi^{(i)}=&[\xi^q_{in,1},\xi^p_{in,1},\xi^q_{in,2},\xi^p_{in,2}]^T,  ~~\xi^{(o)}=[\xi^q_{out,1},\xi^p_{out,1},\xi^q_{out,2},\xi^p_{out,2}]^T, \nonumber\\
\xi_1=&[\xi^{(o)}, \xi_{in}]^T,  ~~ \xi_2=[\xi^{(i)}, \xi_{out}]^T, \nonumber\\
\xi'=&[\xi_{in}, \xi_{loss}]^T, ~~ \xi=[\xi^{(i)}, \xi_{loss}]^T.
\end{align}
\normalsize
According to (\ref{eq:NOPA-dynamics1}) and (\ref{eq:NOPA-dynamics2}), we have
\small
\begin{eqnarray}
\dot{z}\left(t\right) &=& R z\left(t\right)+ \left[\begin{array}{cc}
-\sqrt{\gamma}I_8 &  -\sqrt{\kappa}I_8 \end{array}\right] \xi'\left(t\right), \label{eq: dual-NOPA-dynamics-1} \\
\xi_{out}\left(t\right) &=& \sqrt{\gamma} z\left(t\right)+\xi_{in}\left(t\right),\label{eq: dual-NOPA-dynamics-2}
\end{eqnarray}
\normalsize
where $R$ is as shown in (\ref{eq:matrices}).

Define 
\small
\begin{eqnarray}
\xi_1\left(t\right)&=& S\xi_2\left(t\right),\label{eq:passive-matrix}
\end{eqnarray}
\normalsize
in which the real unitary matrix $S$ is the quadrature form of $\tilde S$, satisfying the following relations
\small
\begin{align}
S^T S = S  S^T = I_{12},\nonumber \\
SJ_nS^T = J_n, \label{eq: real-unitary}
\end{align}
\normalsize
where
\small
\begin{eqnarray}
J_n= I_6 \otimes \left[ \begin{array}{cc} 0 & 1\\ -1 & 0 \end{array} \right]. \label{eq:Jn} 
\end{eqnarray}
\normalsize
Let
\small
\begin{eqnarray}S=\left[
\begin{array}{cc}
S_{11} & S_{12} \\ S_{21} & S_{22} \label{eq:sub-S}
\end{array}\right],
\end{eqnarray} 
\normalsize
with
\small 
\begin{eqnarray}
S_{11} &=& \left[\begin{array}{cc} I_4 &  O_{4 \times 8} \end{array}\right] S \left[\begin{array}{c} I_4 \\  O_{8 \times 4} \end{array}\right], \nonumber \\
S_{21} &=& \left[\begin{array}{cc} O_{8 \times 4} &  I_8 \end{array}\right] S \left[\begin{array}{c} I_4 \\  O_{8 \times 4} \end{array}\right], \nonumber\\
S_{12} &=& \left[\begin{array}{cc} I_4 &  O_{4 \times 8} \end{array}\right] S \left[\begin{array}{c} O_{4 \times 8} \\ I_8  \end{array}\right],\nonumber \\
S_{22} &=& \left[\begin{array}{cc} O_{8 \times 4} & I_8  \end{array}\right] S \left[\begin{array}{c} O_{4 \times 8} \\ I_8 \end{array}\right]. 
\end{eqnarray}
\normalsize
We have
\small
\begin{eqnarray}
\xi^{(o)} &=& S_{11}\xi^{(i)}+S_{12}\xi_{out}, \label{eq: eq-1}\\
\xi_{in} &=& S_{21}\xi^{(i)}+S_{22}\xi_{out} \label{eq: eq-2}. 
\end{eqnarray}
\normalsize
Based on (\ref{eq: dual-NOPA-dynamics-1}), (\ref{eq: dual-NOPA-dynamics-2}), (\ref{eq: eq-1}) and (\ref{eq: eq-2}), we obtain (\ref{eq: system-dynamics}).

\section*{APPENDIX 2}
$\tilde{S}_{lm}$ and its corresponding gradient $G=D_{\tilde{S}}$ are as shown in (\ref{eq:S-optimized}) and (\ref{eq:G-optimized}), respectively.
\begin{figure*}[!b]
\scalebox{0.6}{%
 \begin{minipage}{1.0\linewidth}
 \begin{eqnarray}
\tilde{S}_{lm} =\left[\begin{array}{cccccc} -0.012305658659326  & 0.000000000000071  & 0.008576364236157 & -0.000000000000142 &  0.999887502042829 &  0.000000000000110\\
  -0.000000000000071 & -0.012305658659326 & -0.000000000000109 &  0.999887502042830  & 0.000000000000142 &  0.008576364236157\\
   0.999887502042829  & 0.000000000000001 & -0.008471156255372 &  0.000000000000069  & 0.012378318554964 & -0.000000000000048\\
   0.000000000000051 &  0.008576364236158  & 0.000000000000085 & -0.008471156255372 & -0.000000000000112 &  0.999927340104363\\
   0.008576364236157 & -0.000000000000050 &  0.999927340104363 &  0.000000000000111 & -0.008471156255372 & -0.000000000000085\\
  -0.000000000000001 &  0.999887502042829 &  0.000000000000047  & 0.012378318554963 & -0.000000000000069 &  -0.008471156255373
\end{array}\right]\label{eq:S-optimized}
\end{eqnarray} 
 \end{minipage}
}
\end{figure*}
\begin{figure*}[!t]
\scalebox{0.6}{%
 \begin{minipage}{1.0\linewidth}
 \begin{eqnarray}
G =10^{-3}\left[\begin{array}{cccccc} 0.409017637139186 & -0.000013389581000 &-0.292752071754183 & -0.131726307590905 & 0.007542180013146 & 0.092604228142405\\
0.000013384124031 & 0.409017646234133 & -0.092604224236176 & 0.007631674074983 & 0.131726307489625 & -0.292814986508976\\
0.007586926959628 & -0.000000246987169 & -0.005430300582194 & -0.002443410193280 & 0.000139901426214 & 0.001717728155075\\
-0.000009579672002 & -0.292783532386752 & 0.066288073024820 & -0.005462914965842 & -0.094292493155211 & 0.209603196744500\\
-0.292783525876746 & 0.000009583578405 & 0.209558161024828 & 0.094292493228142 & -0.005398853213792 & -0.066288075820747\\
0.000000246886333 & 0.007586927128612 & -0.001717728084123 & 0.000141561463910 & 0.002443410191339 & -0.005431467597355
\end{array}\right] \label{eq:G-optimized}
\end{eqnarray} 
 \end{minipage}
}
\end{figure*}
The group of two-level unitary matrices $\tilde{S}_{k}$ ($k \in [1,13]$)  as the decomposition of $ \tilde{S}_{lm}$ in Section~\ref{sec:decomposition } is
\small
\begin{eqnarray}
&&\tilde{S}_1= \tilde{S}_{15}= I_6,\nonumber \\
&&\tilde{S}_2=\tilde{S}_{7}=\tilde{S}_{11}=\tilde{S}_{14}=\left[\begin{array}{cccccc} 1 & 0 & 0 & 0 & 0 & 0 \\
0 & 0 & 1 & 0 & 0 & 0 \\ 0 & -1 & 0 & 0 & 0 & 0 \\
0 & 0 & 0 & 1 & 0 & 0 \\ 0 & 0 & 0 & 0 & 1 & 0 \\  0 & 0 & 0 & 0 & 0 & 1 
\end{array}\right],\nonumber
~~ \tilde{S}_3 = \left[\begin{array}{cccccc} 1 & 0 & 0 & 0 & 0 & 0 \\
0 & 1 & 0 & 0 & 0 & 0 \\ 0 & 0 & \alpha_3 & \beta_3 & 0 & 0 \\
0 & 0 & -\beta_3 & \alpha_3 & 0 & 0 \\ 0 & 0 & 0 & 0 & 1 & 0 \\  0 & 0 & 0 & 0 & 0 & 1 
\end{array}\right], \\
&&\tilde{S}_4 =\tilde{S}_9 = \left[\begin{array}{cccccc} 1 & 0 & 0 & 0 & 0 & 0 \\
0 & 1 & 0 & 0 & 0 & 0 \\ 0 & 0 & 1 & 0 & 0 & 0 \\
0 & 0 & 0 & 0 & 1 & 0 \\ 0 & 0 & 0 & -1 & 0 & 0 \\  0 & 0 & 0 & 0 & 0 & 1 
\end{array}\right],
~~~~~~~~~~~~~~~~~~~\tilde{S}_5 =\left[\begin{array}{cccccc} 1 & 0 & 0 & 0 & 0 & 0 \\
0 & 1 & 0 & 0 & 0 & 0 \\ 0 & 0 & 1 & 0 & 0 & 0 \\
0 & 0 & 0 & 1 & 0 & 0 \\ 0 & 0 & 0 & 0 & \alpha_5 & \beta_5 \\  0 & 0 & 0 & 0 & -\beta_5 & \alpha_5 
\end{array}\right],\nonumber\\
&&\tilde{S}_6 = \left[\begin{array}{cccccc} \alpha_6 & \beta_6 & 0 & 0 & 0 & 0 \\
-\beta_6 & \alpha_6 & 0 & 0 & 0 & 0 \\ 0 & 0 & 1 & 0 & 0 & 0 \\
0 & 0 & 0 & 1 & 0 & 0 \\ 0 & 0 & 0 & 0 & 1 & 0 \\  0 & 0 & 0 & 0 & 0 & 1 
\end{array}\right],
~~~~~~~~~~~~~~~~~~~~~~~\tilde{S}_8 = \left[\begin{array}{cccccc} 1 & 0 & 0 & 0 & 0 & 0 \\
0 & 1 & 0 & 0 & 0 & 0 \\ 0 & 0 & \alpha_8 & \beta_8 & 0 & 0 \\
0 & 0 & -\beta_8 & \alpha_8 & 0 & 0 \\ 0 & 0 & 0 & 0 & 1 & 0 \\  0 & 0 & 0 & 0 & 0 & 1 
\end{array}\right],\nonumber\\
&&\tilde{S}_{10} = \left[\begin{array}{cccccc} -1 & 0 & 0 & 0 & 0 & 0 \\
0 & -1 & 0 & 0 & 0 & 0 \\ 0 & 0 & 1 & 0 & 0 & 0 \\
0 & 0 & 0 & 1 & 0 & 0 \\ 0 & 0 & 0 & 0 & 1 & 0 \\  0 & 0 & 0 & 0 & 0 & 1 
\end{array}\right],
~~~~~~~~~~~~~~~~~~~~~~\tilde{S}_{12} = \left[\begin{array}{cccccc} 1 & 0 & 0 & 0 & 0 & 0 \\
0 & 1 & 0 & 0 & 0 & 0 \\ 0 & 0 & \alpha_{12} & \beta_{12} & 0 & 0 \\
0 & 0 & -\beta_{12} & \alpha_{12} & 0 & 0 \\ 0 & 0 & 0 & 0 & 1 & 0 \\  0 & 0 & 0 & 0 & 0 & 1 
\end{array}\right],\nonumber\\
&&\tilde{S}_{13} = \left[\begin{array}{cccccc} \alpha_{13} & \beta_{13} & 0 & 0 & 0 & 0 \\
-\beta_{13} & \alpha_{13} & 0 & 0 & 0 & 0 \\ 0 & 0 & 1 & 0 & 0 & 0 \\
0 & 0 & 0 & 1 & 0 & 0 \\ 0 & 0 & 0 & 0 & 1 & 0 \\  0 & 0 & 0 & 0 & 0 & 1 
\end{array}\right]. \label{eq:group_S_i}
\end{eqnarray}

\end{document}